\documentclass[aps,prc,superscriptaddress, preprintnumbers, nofootinbib,10pt, showpacs, floatfix]{revtex4-1}

\usepackage{hyperref}

\usepackage{multirow}

\usepackage{feynmf}

\usepackage{amsmath,amssymb}

\usepackage{graphicx,bm}

\usepackage{slashed}

\usepackage{natbib}

\usepackage{epstopdf}

\usepackage{ulem} 

\usepackage[usenames]{color}

\usepackage{float}

\renewcommand\sout{\bgroup \color{red} \ULdepth=-.5ex \ULset}

\makeatletter
\newcommand\makebig[2]{%
  \@xp\newcommand\@xp*\csname#1\endcsname{\bBigg@{#2}}%
  \@xp\newcommand\@xp*\csname#1l\endcsname{\@xp\mathopen\csname#1\endcsname}%
  \@xp\newcommand\@xp*\csname#1r\endcsname{\@xp\mathclose\csname#1\endcsname}%
}
\makeatother

\makebig{Biggg} {3.0}
 
\begin{document}
\title{Excluded-volume model for quarkyonic Matter II: Three-flavor shell-like distribution of baryons in phase space}

\author{Dyana C. Duarte}
\email[]{dyduarte@uw.edu}
\author{Saul Hernandez-Ortiz}
\email[]{saulhdz@uw.edu}
\author{Kie Sang Jeong}
\email[]{ksjeong@uw.edu}
\affiliation{Institute for Nuclear Theory, University of Washington, Seattle, WA 98195, USA}

\date{\today}

\preprint{INT-PUB-20-028}

\begin{abstract}
We extend the excluded-volume model of isospin symmetric two-flavor dense quarkyonic matter [Phys. Rev. C \textbf{101}, 035201 (2020)] including strange particles  
and address its implications for neutron stars. The effective sizes of baryons are defined from the diverging hard-core potentials in the short interdistance regime. Around the hard-core density, the repulsive core  between baryons at short distances leads to a saturation in the number density of baryons and generates  perturbative quarks from the lower phase space, which leads to the shell-like distribution of baryons  by the Pauli exclusion principle. The strange-quark Fermi sea always appears at high densities but the $\Lambda$ hyperon shell only appears when the effective size of the $\Lambda$ hyperon is smaller than the effective size of nucleons. We find that the pressure of strange quarkyonic matter can be large enough to support neutron stars with two times solar mass and can have a large sound speed, $c_s^2 \simeq 0.7$. The fraction of the baryon number carried by perturbative quarks is about 30\% at the inner core of most massive neutron stars. 
\end{abstract}
\maketitle

\section{Introduction}\label{intro}

Observation of GW170817~\cite{TheLIGOScientific:2017qsa, Abbott:2018exr} provided important information for understanding dense nuclear matter. The possible range of tidal deformability is confined with 90\% confidence level~\cite{TheLIGOScientific:2017qsa, Abbott:2018exr} and the subsequent analyses constrained the corresponding radius $R_{1.4}\leq 13.5~\textrm{km}$~\cite{Fattoyev:2017jql, Annala:2017llu, Vuorinen:2018qzx, Raithel:2018ncd, Most:2018hfd, Tews:2019cap, Tews:2019ioa, Capano:2019eae}. A recent GW190425 observation~\cite{Abbott:2020uma} constrained the possible range $R_{M\geq1.4} \leq 15~\textrm{km}$, including higher mass states. Meanwhile, an equation of state (EoS) hard enough to support a twice
solar mass ($M_{\odot}$) state is required, which normally leads to larger radius star~\cite{Demorest:2010bx, Antoniadis:2013pzd, Cromartie:2019kug}. To reconcile these observations, the EoS should be soft enough in the low density regime and hard enough in the high density regime so that the strong pressure of the inner core at higher densities can support a larger mass state and the weaker pressure of the outer core at lower densities can satisfy the $R_{1.4}\leq 13.5~\textrm{km}$ constraint. Then, the expected soft-hard evolution of the EoS should accompany the sound velocity $c_s^2 > 0.3$ around a few times the normal nuclear density ($\rho_0$)~\cite{Masuda:2012ed, Alford:2013aca, Kojo:2014rca, Bedaque:2014sqa,Tews:2018kmu, Ma:2018qkg, Greif:2018njt, Fujimoto:2019hxv, Kojo:2019raj}. Beyond the density regime of the inner core where a hard EoS is supported, a softened EoS is expected under the causality and conformal limit constraints~\cite{Annala:2017llu, Vuorinen:2018qzx, Kojo:2014rca, Bedaque:2014sqa,Tews:2018kmu, Ma:2018qkg, Greif:2018njt, Fujimoto:2019hxv, Kojo:2019raj, Annala:2019puf}.

However, it is hard to reconcile both constraints from fundamental principles. If one considers mean-field potentials between baryons, certain universal repulsive contributions are expected for the EoS~\cite{Hebeler:2013nza, Gandolfi:2013baa} at high densities as the newly generated degrees of freedom lead to a soft EoS through various decay channels into the low energy states~\cite{Glendenning:1984jr, Knorren:1995ds, Brown:1975di, Cai:2015hya}. Even if  stiff evolution is obtained by some kind of model, it is hard to explain the expected softening evolution at the high density limit by these same first principles. Some kind of phase transition to quark matter can be introduced.  A phase transition to quark matter  attenuates the hard nature of the  EoS.  There is much literature that discussing  the signals of such a hypothetical phase transition~\cite{Masuda:2012ed, Alford:2013aca, Kojo:2014rca, Benic:2015pia, Marczenko:2017huu, Annala:2019puf, Marczenko:2020jma, Marczenko:2020wlc, Ma:2020hno}. As an alternative  candidate for a solution, it is worthwhile  to consider a quarkyonic-like model~\cite{McLerran:2007qj, Fukushima:2015bda, McLerran:2018hbz, Jeong:2019lhv, Duarte:2020xsp, Zhao:2020dvu} which naturally generates the hard-soft evolution of the EoS. 

Quarkyonic matter is based on large-$N_c$ quantum chromodynamics (QCD)~\cite{tHooft:1973alw, tHooft:1974pnl}. In this model, the quasiquark states on the surface of the large Fermi sea are confined into baryon-like confined states as the quark confinement mechanism of the baryon state in vacuum is not altered by the hard scale of the quark chemical potential. Therefore, one can expect a shell-like distribution of the baryons ($k_{F}^{b}\geq N_c k_{F}^{Q}$) and the quasifree quarks occupying the lower phase sphere by the Pauli exclusion principle. Once the lower phase space is saturated by the quarks, a rapid enhancement of chemical potential of the baryon-like state  [$k_{F}^{b}\sim O( \Lambda_{\textrm{QCD}}) \rightarrow N_c k_{F}^{Q}$] is expected, which leads to the required stiff evolution of the EoS. This is not a first-order phase transition because the pressure is not fixed but increases suddenly and smoothly by the enhanced chemical potential and there is no discontinuity for the increment of energy density and the baryon number density~\cite{McLerran:2007qj}. At the extremely high density limit, perturbative QCD matter will appear as the Debye screening begins to block the confinement process [$r_{\textrm{Debye}} \sim O({N_c}^{0})$].

This concept was introduced to describe the hard-soft evolution of the EoS in  previous literature~\cite{McLerran:2018hbz, Jeong:2019lhv, Duarte:2020xsp, Zhao:2020dvu}. Model construction with an explicit shell-like distribution~\cite{McLerran:2018hbz} produced a plausible result satisfying the  aforementioned constraints, and a two-flavor generalization was studied under the $\beta$-equilibrium condition~\cite{Zhao:2020dvu}. In the phenomenological model construction, one can consider the hard-core repulsive interaction whose scale can be regarded as the effective size of the baryon~\cite{Hamada:1962nq, Herndon:1967zza, Carnahan:1969, Bethe:1971xm, Kurihara:1984mh, Rischke:1991ke, Kievsky:1992um, Stoks:1994wp, Wiringa:1994wb, Yen:1997rv, Machleidt:2000ge, Vovchenko:2015vxa, Zalewski:2015yea,  Redlich:2016dpb, Alba:2016hwx, Vovchenko:2017cbu, Vovchenko:2017zpj, Motornenko:2019arp, Lourenco:2019ist, Dutra:2020qsn}. In the isospin symmetric excluded-volume model~\cite{Jeong:2019lhv}, the repulsive core dynamically generates the shell-like phase structure of baryons which reproduces the stiff evolution of the EoS with $c_s^2\simeq 0.7$ as analyzed in the literature~\cite{Fattoyev:2017jql, Annala:2017llu, Vuorinen:2018qzx, Raithel:2018ncd, Most:2018hfd,  Tews:2019cap, Tews:2019ioa, Capano:2019eae, Masuda:2012ed, Kojo:2014rca, Bedaque:2014sqa, Ma:2018qkg, Tews:2018kmu, Kojo:2019raj, Greif:2018njt, Fujimoto:2019hxv, Kojo:2019raj}. In the previous work~\cite{Duarte:2020xsp}, the excluded-volume model was extended to the three-flavor mixture of baryon and quarks where the scale of the repulsive core is adopted from first-principles studies~\cite{Ishii:2006ec, Inoue:2016qxt, Nemura:2017bbw, Hatsuda:2018nes, Inoue:2018axd,  Sasaki:2019qnh, Park:2018ukx, Park:2019bsz}. We argued in favor of the dynamical role of the multi flavor hard-core repulsion in Ref.~\cite{Duarte:2020xsp}, but, as the shell-like distribution was omitted, the resulting EoS was not hard enough to satisfy the physical constraints.

In this paper, we present the three-flavor excluded-volume model with an explicit shell-like distribution of baryons appearing after the saturation momentum of the quark Fermi sea. The paper is organized as follows. In Sec.~\ref{sec2}, we present a brief introduction of the excluded-volume model and the possible structure of the shell-like distribution. In Sec.~\ref{sec3}, we explain the physical configurations, EoS, and corresponding mass-radius relations obtained under the physical constraints including the electromagnetic charge neutrality and the equilibrium constraints from weak interactions. Finally, in Sec.~\ref{sec4} we summarize our results and discuss possible developments for the future work.

\section{Configuration of three-flavor quarkyonic matter}\label{sec2}

In dense quarkyonic matter~\cite{McLerran:2007qj}, the quark wave functions distributed around the quark Fermi surface are clearly confined in baryon-like states because Debye screening is suppressed in the large-$N_c$ limit~\cite{tHooft:1974pnl, tHooft:1973alw}. The matter looks like normal nuclear matter in the low density regime as the momentum of a quark is distributed in the confinement range. However, when the matter density reaches few times $\rho_0$, where $k_F^{b} \sim O(\Lambda_{\textrm{QCD}}$) so that lowest momentum states become distributed away from the clear confinement range, the quark Fermi sea is formed from the low momentum phase space. When the lower momentum space is saturated by quasifree quarks, the confined quark momenta should be larger than the Fermi momentum of the saturated quarks by the Pauli exclusion principle, which leads to the shell-like momentum distribution of baryons~\cite{McLerran:2007qj, McLerran:2018hbz, Jeong:2019lhv}. Around the onset moment, the pressure of the  system will be continuously and stiffly increasing as the evolution of chemical potential should show stiffness and continuity ($k_{F}^{b} \simeq N_c k_{F}^{Q}$), contrary to the expected evolution in a first-order phase transition. The phenomenological configuration strongly depends on how one defines the lower boundary of the distribution because the dynamical equilibrium constraints are related through the shell-like distribution. In this section, we will briefly introduce the excluded-volume model approach, and present the explicit structure of the shell-like baryon distribution formed by the dynamically saturated quark Fermi sea. We will use the following abbreviations to denote the baryons and quarks:  $B$ represents the total baryons including quarks, $b$ represents the baryon (hadron), and $Q$ represents the saturated quarks; $b_i$ represents the baryon flavors \{$n,p,\Lambda$\} and $Q_i$ represents the quark flavors \{$u,d,s$\}.

\subsection{Brief summary of the excluded-volume model for quarkyonic-like matter}\label{sec2a}

As introduced in the previous literature~\cite{Jeong:2019lhv, Duarte:2020xsp}, one can simplify the baryon-baryon central potential whose strong repulsive core is expected at high density regime~\cite{Inoue:2016qxt, Nemura:2017bbw, Hatsuda:2018nes, Inoue:2018axd, Park:2018ukx, Park:2019bsz, Sasaki:2019qnh} by supposing an infinite-well shaped potential whose hard-core radius is around $r_c \simeq 0.6~\textrm{fm}$ scale. Among the low-lying baryon octet, nucleons and the $\Lambda$ hyperon would be the lightest particles which have a strong repulsive core at short interdistance according to the lattice QCD calculation~\cite{Inoue:2016qxt, Nemura:2017bbw, Hatsuda:2018nes, Inoue:2018axd, Sasaki:2019qnh}. Thus, in the dense regime, a three-flavor baryon ($n$, $p$, and $\Lambda$) system can be suggested as a simplest multiflavor extension where the effective size of particle is understood from the hard-core repulsion around  $n_B \sim n_0$~\cite{Inoue:2016qxt, Nemura:2017bbw, Hatsuda:2018nes, Inoue:2018axd, Park:2018ukx, Park:2019bsz, Sasaki:2019qnh}. If only the quasibaryons are assumed, the number density in the excluded volume can be defined as follows~\cite{Jeong:2019lhv, Duarte:2020xsp}:
\begin{align}
  n_{b_i}^{ex} &= \frac{n_{b_i}}{1 - \tilde{n}_{b}/n_0}=\frac{2}{(2\pi)^3} \int^{K_{F}^{b_i}}_{0}d^3 k,\label{exd}\\
  \tilde{n}_{b} &= n_{n}+n_{p}+(1+\alpha) n_{\Lambda},\label{baa}
\end{align}
where $K_{F}^{b_i}$ represents the enhanced Fermi momentum due to the reduced available volume and $\alpha$ determines the strength of the hard core repulsive interaction between the surrounding baryon and the $\Lambda$ hyperon in the range of $\vert \alpha \vert<0.2$.\footnote{ In the context of the presumed effective size of a particle, this approach could be understood as the cold-dense limit of the van der Waals (vdW) EoS in Fermi-Dirac statistics~\cite{Rischke:1991ke, Kievsky:1992um, Stoks:1994wp,  Wiringa:1994wb, Yen:1997rv, Machleidt:2000ge, Vovchenko:2015vxa, Zalewski:2015yea, Redlich:2016dpb, Alba:2016hwx, Vovchenko:2017cbu, Vovchenko:2017zpj, Motornenko:2019arp, Lourenco:2019ist, Dutra:2020qsn}. As a simple example, $K^b_F$ can be obtained from $\mu^{*}=\mu^{\textrm{id}}(n_b^{ex},T\rightarrow0)$ without an attraction term if one derives the intensive number from the vdW EoS in Fermi-Dirac statistics~\cite{Vovchenko:2015vxa}. } However, as we focus on the high density regime where the interdistance of particles becomes of the order of the hard-core radius, $n_0 > 0.65~\textrm{fm}^{-3} \simeq 4\rho_0$ will be considered, which is a different order of magnitude from the size used in Refs.~\cite{Vovchenko:2015vxa, Zalewski:2015yea, Redlich:2016dpb, Alba:2016hwx, Vovchenko:2017cbu, Vovchenko:2017zpj, Motornenko:2019arp, Lourenco:2019ist, Dutra:2020qsn}. One may adopt a well constructed model~\cite{Hebeler:2013nza, Gandolfi:2013baa, Motornenko:2019arp, Kojo:2014rca, Tews:2018kmu, Kojo:2019raj} and use a Maxwell construction to accommodate the low density properties of nuclear matter. The variation range $\vert \alpha \vert<0.2$ for the hard-core size of $\Lambda$  is supposed by considering the possible error band of the $\Lambda$N potential from lattice QCD~\cite{Inoue:2016qxt, Nemura:2017bbw, Hatsuda:2018nes, Inoue:2018axd}, which is relatively small in comparison with the variation range studied in Refs.~\cite{Alba:2016hwx, Vovchenko:2017zpj}. If the SU(3) flavor symmetry breaking term is non-negligible or kaon condensation plays a significant role~\cite{Kaplan:1986yq, Savage:1995kv, Jeong:2016qlk}, the effective size of $\Lambda$ can be different from the current range of variation.

The energy density of the corresponding system can be described as the one of non-ideal free fermions having effective size~\cite{Jeong:2019lhv, Duarte:2020xsp}:
\begin{align}
\varepsilon_{b} & = \left( 1-  \frac{\tilde{n}_{b}}{n_0} \right) \frac{1}{\pi^2} \sum_{i}^{\{n,p,\Lambda\} } \int^{K_{F}^{b_i}}_{0} dk k^2  \left(k^2 + m_{b_i}^2\right)^{\frac{1}{2}}+ \frac{(3 \pi^2)^{\frac{4}{3}}}{4\pi^2}n_{e}^{\frac{4}{3}},
\end{align}
where the electron mass is assumed to be small compared to the Fermi momentum scale. If one takes the nonrelativistic limit, the baryon chemical potential can be obtained as follows:
\begin{align}
\mu_i   \simeq &~m_{b_i} +  \frac{ (3 \pi^2)^{\frac{5}{3}}}{10 \pi^2 m_{i} } \frac{5}{3} {n_{b_i}^{ex}}^{\frac{2}{3}} + \omega_{i}  \sum_{j  }^{\{n,p,\Lambda\}} \frac{ (3 \pi^2)^{\frac{5}{3}}}{10 \pi^2 m_{j}} \frac{2}{3 n_0} {n_{b_{j}}^{ex}}^{\frac{5}{3}}+\cdots,\label{hchemp}
\end{align}
where $\omega_{i}=\partial \tilde{n}_{b}  / \partial n_{i}$ ($\omega_{n,p}=1$, $\omega_{\Lambda}=1+\alpha$). As one can find from the third term, the chemical potential of a specific flavor~\eqref{hchemp} can be enhanced without having large ${n_{b_i}^{ex}}$ if the some part of system volume is occupied by the other finite size particles. Thus, to accommodate a heavier baryon (denoted flavor $h$), its effective size should be small so that the contribution from the third term is suppressed ($\omega_h \ll 1$). Therefore, it is unlikely to have the higher mass state such as $\Delta(1232)$ if the particle has an effective size of order similar to $n_0$.  Due to the intrinsic divergence around $n_B \sim n_0$, this system cannot accommodate $n_B>n_0$ and contains an unphysical configuration ($v_s^2 \gg 1$). 

If new degrees of freedom are considered, as is done in the Hagedorn model~\cite{Hagedorn:1965st}, the dynamically generated quark degrees freedom lead to a physically plausible explanation in accordance with the quarkyonic matter concept~\cite{Jeong:2019lhv, Duarte:2020xsp}. Once the quark Fermi sea is saturated, the baryons should have the shell-like distribution in  momentum space as a consequence of the Pauli exclusion principle: the quarks confined in the baryon should have momentum larger than the saturated quark Fermi momentum.  If one assumes the isospin symmetric quark configuration as  discussed in Ref.~\cite{Jeong:2019lhv}, the lower boundary of the distribution is simply obtained as $k_{F}^{b}=N_c k_{F}^Q$. Even if the asymmetric configuration is considered, the scale can be estimated to be around $k_{F}^{b} \sim N_c \textrm{max.} \left[ k_{F}^{u}, k_{F}^{d}, k_{F}^{s}  \right]$ as the quarks confined in a baryon should have a common scale of momentum. A detailed argument for the definition of ${k_{F}^{b_i}}$ in the isospin asymmetric configuration will be given in the next subsection.  The baryon number in the excluded-volume density within the explicit shell-like structure can be written as
\begin{align}
  \bar{n}_{b_i}^{ex} &= \frac{n_{b_i}}{1 - \tilde{n}_{b}/n_0}=\frac{2}{(2\pi)^3} \int^{\left[ k_{F} + \Delta \right]_{b_i}}_{k_{F}^{b_i}}d^3 k,\label{exds}
\end{align}
where the upper boundary of the baryon distribution has been defined by assuming the fully occupied phase space:
\begin{align}
 \left[k_{F} + \Delta \right]_{b_i} & = \left(3\pi^2  \bar{n}_{b_i}^{ex} + {k_{F}^{b_i}}^3 \right)^{\frac{1}{3}},\label{b-q1}
\end{align}
where the $\Delta$ is the width of the baryon distribution~\cite{Jeong:2019lhv}.

In quarkyonic matter, the quark Fermi sea would be continuously saturated without any signature of a  first-order phase transition according to large-$N_c$ gauge dynamics~\cite{McLerran:2007qj, Jeong:2019lhv, Duarte:2020xsp}. Thus, smooth interpolation of a quark's energy should be possible in both directions around the Fermi surface. In this excluded volume model approach, the energy interpolation is continuous by analytic definition, but an  unphysical divergence appears at the onset moment of saturation. A Huge energy enhancement due to the sudden formation of the shell-like baryon distribution leads to the unphysical energy dispersion relation corresponding to $\partial n_{B}/\partial n_{\tilde{Q}} \gg 1$, $\partial n_{B}/\partial n_{b} \ll 0$ ($n_B=n_{b}+n_{\tilde{Q}}$ where the tilde in the subscript denotes the number density in baryon units). To attenuate the unphysical divergence, an enhanced phase measure $\mathcal{M}_{i}(k^2)$ for the saturated quarks can be introduced. The modified measure $\mathcal{M}_{i}(k^2) > k^2$ effectively enhances the free quark density around the saturation moment of free quarks and converges to the ideal gas limit [$\mathcal{M}_{i}(k^2) \rightarrow  k^2$] in the high density regime ($k_F^{Q_i} \gg \Lambda_{\textrm{QCD}} $). Then the quark number density can be written in baryon number units as follows:
\begin{align}
n_{\tilde{Q}_{i}}\equiv \frac{1}{\pi^2} \int^{k_F^{Q_{i}}}_{0} dk \mathcal{M}_{i}(k^2).
\end{align}
The relatively rapid growth of quark density at the onset moment ($n_{b} \simeq n_0$) makes an effective barrier for $\delta n_{\tilde{Q}}$ in the variation of the baryon number density ($ n_{b} < n_{0} - \delta n_{\tilde{Q}} $), which prevents the unphysical divergence and leads to the gradual formation of the shell-like distribution. The energy density with explicit shell-like baryon distribution can be written as follows:
\begin{align}
\varepsilon_{\textrm{qy.}} &= 2 \left( 1- \frac{  \tilde{n}_{b}}{n_0} \right) \sum_i^{\{n,p,\Lambda\}} \int_{k^{b_i}_F}^{\left[k_F+\Delta\right]_{b_i}} \frac{d^3 k}{(2\pi)^3} \left(k^2 + m_{b_i}^2 \right)^{\frac{1}{2}}+ \frac{ N_c}{\pi^2} \sum_j^{\{u,d,s\}} \int^{k^{Q_j}_F}_0  d k \mathcal{M}_{j}(k^2)\left(k^2 + m_{Q_j}^2\right)^{\frac{1}{2}}+ \frac{(3 \pi^2)^{\frac{4}{3}}}{4\pi^2}n_{e}^{\frac{4}{3}}.
\end{align}
The corresponding baryon ($n,p$, and $\Lambda$) chemical potential can be obtained as
\begin{align}
\mu_{b_i} = \frac{\partial \varepsilon_{\textrm{qy.}}}{\partial n_{b_i}} = &\left( 1- \frac{ \tilde{n}_{b}}{n_0} \right) \Biggg\{ \frac{  \left[k_F+\Delta\right]_{b_i}^2}{\pi^2} \left(\left[k_F+\Delta\right]_{b_i}^2 + m_{b_i}^2 \right)^{\frac{1}{2}} \frac{\partial  \left[k_F+\Delta\right]_{b_i}}{\partial n_{b_i} }  \nonumber\\
&\qquad\qquad\qquad +   \sum_{j \neq i}^{\{n,p,\Lambda\}}  \frac{\left[k_F+\Delta\right]_{b_j}^2}{\pi^2} \left(\left[k_F+\Delta\right]_{b_j}^2 + m_{b_j}^2 \right)^{\frac{1}{2}} \frac{\partial \left[k_F+\Delta\right]_{b_j} }{\partial n_{b_i} } \Biggg\} \nonumber\\
&\qquad\qquad\qquad\quad-\frac{\omega_i}{n_0}\sum_k^{\{n,p,\Lambda\}} \frac{1}{\pi^2} \int_{k^{b_k}_F}^{\left[k_F+\Delta\right]_{b_k}} dk k^2  \left(k^2 + m_{b_k}^2 \right)^{\frac{1}{2}},\nonumber\\
=&  \left(  \frac{n_0 -(\tilde{n}_{b}-\omega_i n_{b_i})}{n_0 -\tilde{n}_{b}} \right) \left(\left[k_F+\Delta\right]_{b_i}^2 + m_{b_i}^2 \right)^{\frac{1}{2}}  \nonumber\\
&~+  \frac{\omega_i}{n_0}\left\{ \sum_{j \neq i}^{\{n,p,\Lambda\}}   \bar{n}_{b_j}^{ex}\left(\left[k_F+\Delta \right]_{b_j}^2 + m_{b_j}^2 \right)^{\frac{1}{2}} -\sum_k^{\{n,p,\Lambda\}} \frac{1}{\pi^2} \int_{k^{b_k}_F}^{\left[k_F+\Delta\right]_{b_k}} dk k^2  \left(k^2+m_{b_k}^2 \right)^{\frac{1}{2}} \right\},\label{bchmp}
\end{align}
where the partial derivatives are calculated as
\begin{align}
\frac{\partial \left[k_F+\Delta\right]_{b_i} }{\partial n_{b_i} }=&~\frac{\pi^2}{\left[k_F+\Delta\right]_{b_i}^2}  \left( \frac{1}{1-\tilde{n}_{b}/n_0 }\right)^{2}\left( 1-  \frac{\tilde{n}_{b}-\omega_i n_{b_i}}{n_0} \right),\\
\frac{\partial \left[k_F+\Delta\right]_{b_j} }{\partial n_{b_i} }=&~\frac{\pi^2}{\left[k_F+\Delta\right]_{b_j}^2} \left( \frac{1}{1-\tilde{n}_{b}/n_0 }\right)^{2} \left( \frac{\omega_{i}  n_{b_j}}{n_0} \right),
\end{align}
with $\omega_{n,p}=1$, $\omega_{\Lambda}=1+\alpha$. Again, the characteristic feature of the excluded-volume model can be found from the $\omega_i$ dependent terms of the chemical potential~\eqref{bchmp}. Even if there exist only a few specific flavors of baryon, the corresponding chemical potential can be enhanced if the space is taken by the other baryons. By the same reasoning, we only consider three-flavors for the baryon side as $n$,~$p$, and $\Lambda$ are expected to have similar orders of $n_0$ \footnote{A detailed argument for the possible emergence of $\Delta(1232)$ is given in the Appendix~\ref{appxa}.}. The quark chemical potential in baryon units can be obtained in a similar way:
\begin{align}
\mu_{\tilde{Q}_i} =&~ \frac{\partial \varepsilon_{\textrm{qy.}}}{\partial n_{\tilde{Q}_i}}\nonumber\\
 = &\left( 1- \frac{  \tilde{n}_{b}}{n_0} \right) \sum_k^{\{u,d,s\}} \left\{ \frac{ \left[k_F+\Delta\right]_{b_k}^2}{\pi^2} \left(\left[k_F+\Delta\right]_{b_k}^2 + m_{b_k}^2 \right)^{\frac{1}{2}} \frac{\partial \left[k_F+\Delta\right]_{b_k} }{\partial  n_{\tilde{Q}_i} }  - \frac{  { k^{b_k}_F }^2}{\pi^2} \left( {k^{b_k}_F}^2 + m_{b_k}^2 \right)^{\frac{1}{2}} \frac{\partial  k^{b_k}_F  }{\partial n_{\tilde{Q}_i} }   \right\} \nonumber\\
&~+ N_c \left(m_{Q_i}^2+\left( k^{Q_i}_F \right)^2 \right)^{\frac{1}{2}}\nonumber\\
=&\left( 1- \frac{  \tilde{n}_{b}}{n_0} \right) \sum_k^{\{u,d,s\}} \frac{\partial {k_{F}^{b_k}}}{ \partial k_F^{Q_{i}}}\frac{{k^{b_k}_F}^2}{ \mathcal{M}_{i}\left({k_F^{Q_{i}}}^2 \right)   }    \left\{ \left( \left[k_F+\Delta\right]_{b_k}^2 + m_{b_k}^2 \right)^{\frac{1}{2}}  - \left( {k^{b_k}_F}^2+ m_{b_k}^2 \right)^{\frac{1}{2}}  \right\}+ N_c \left(\left( k^{Q_i}_F \right)^2 + m_{Q_i}^2 \right)^{\frac{1}{2}},\label{qchemp}
\end{align}
where the partial derivatives are calculated as
\begin{align}
\frac{\partial \left[k_F+\Delta\right]_{b_k}}{\partial n_{\tilde{Q}_i} }=&~ \frac{\pi^2}{\left[k_F+\Delta\right]_{b_k}^2} \frac{  {{k_{F}^{b_k}}}^2   }{\mathcal{M}_{i}\left({k_F^{Q_{i}}}^2 \right) }\frac{\partial {k_{F}^{b_k}}}{ \partial k_F^{Q_{i}}},\label{sdq1}\\
\frac{ \partial k_F^{Q_{i}}} {\partial n_{\tilde{Q}_{i}}} =&~ \frac{\pi^2 }{  \mathcal{M}_{i}\left({k_F^{Q_{i}}}^2 \right)   }.\label{sdq2}
\end{align}
As a consequence of the Pauli exclusion principle, the chemical potential of a saturated quark gets contributions from the baryon distribution as well because $k_{F}^{b_k}$ emerges as a consequence of the saturated quark Fermi sea. Also, one can anticipate another singularity possibly arising in the isospin asymmetric configuration. Suppose the shell-like distribution (${k_{F}^{b_k}} > 0$) formed earlier by saturation of the $d$ quark Fermi sea ($k_F^{d}>0$) and the $u$ quark Fermi sea which is about to appear ($k_F^{u} \simeq 0$). Then, the derivatives~\eqref{sdq1} and \eqref{sdq2} may diverge if $\mathcal{M}_{u}\left({k_F^{u}}^2 \right) \rightarrow 0 $ in $k_F^{u}\rightarrow 0$ limit. To avoid the singularity problem, an infrared regulator can be simply introduced as $\mathcal{M}_{i}\left(k^2 \right) = k^2 + \Lambda_{Q_i}^2$ which leads to the following configurations:
\begin{align}
n_{\tilde{Q}_{i}}&=\frac{1}{\pi^2} \int^{k_F^{Q_{i}}}_{0} dk \left(k^2 + \Lambda_{Q_i}^2\right)= \frac{{ k_F^{Q_{i}}}^3}{3 \pi^2} \left(1+3  \left(  \Lambda_{Q_i} / { k_F^{Q_{i}}} \right)^2 \right),\label{qnd}\\
\varepsilon_{\textrm{qy.}} &= 2 \left( 1- \frac{  \tilde{n}_{b}}{n_0} \right) \sum_i^{\{n,p,\Lambda\}} \int_{k^{b_i}_F}^{\left[k_F+\Delta\right]_{b_i}} \frac{d^3 k}{(2\pi)^3} \left(k^2 + m_{b_i}^2 \right)^{\frac{1}{2}}+ \frac{ N_c}{\pi^2} \sum_j^{\{u,d,s\}} \int^{k^{Q_j}_F}_0  d k \left(k^2 + \Lambda_{Q_j}^2\right) \left(k^2 + m_{Q_j}^2\right)^{\frac{1}{2}}+ \frac{(3 \pi^2)^{\frac{4}{3}}}{4\pi^2}n_{e}^{\frac{4}{3}},\label{defed}
\end{align}
where the criteria for $ \mathcal{M}_{i}\left(k^2 \right)$ are satisfied in both limits. The regulator $\Lambda_{Q_i}$ could be understood as an a
priori non-perturbative contribution remaining on the saturated quark Fermi surface.

\subsection{Explicit structure of a shell-like distribution of baryons}\label{sec2b}

\begin{figure}
\includegraphics[width=0.9\textwidth]{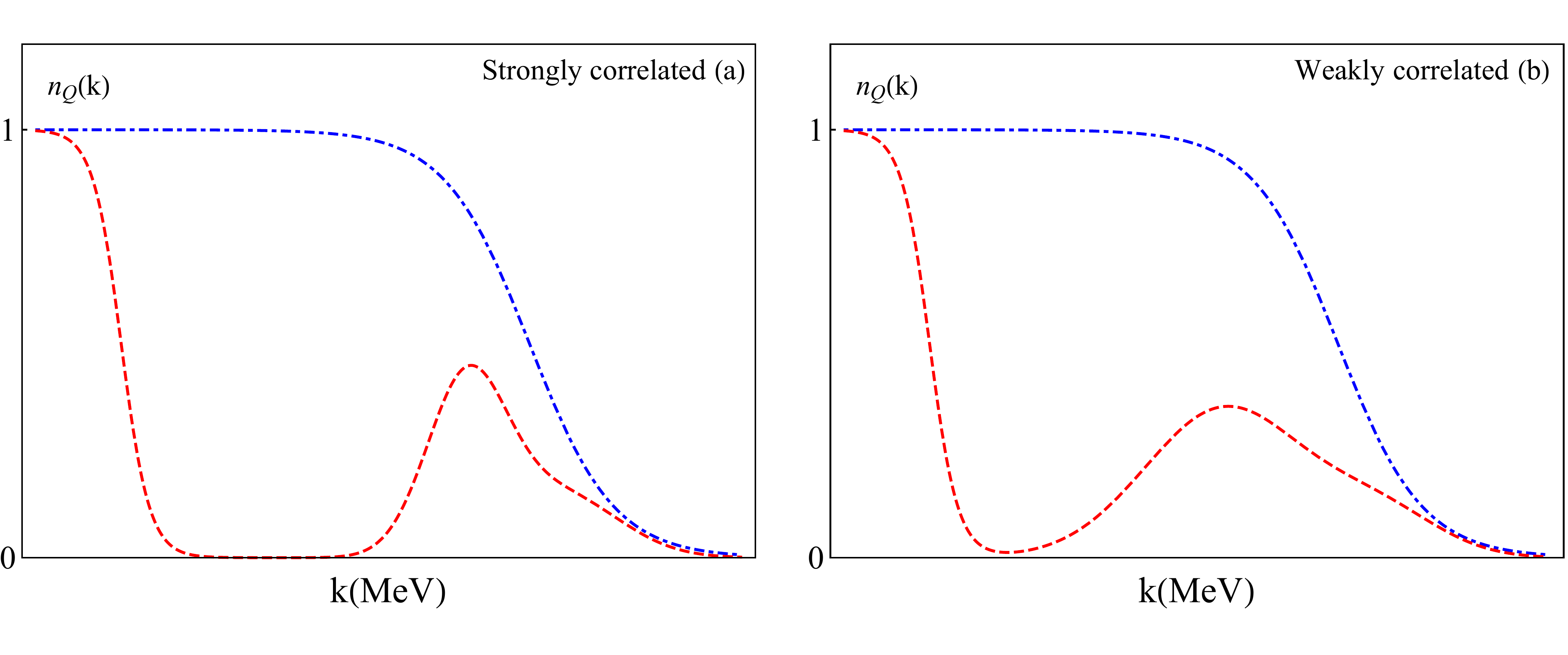} 
\caption{Illustration of quark momentum correlation in the confined state ($k_F^d>k_F^u$). The blue (red) dot-dashed (dashed) line represents the $d$ ($u$) quark distribution. The states whose momentum is distributed in not the fully occupied phase [$n_Q(k)<1$] are understood as confined quark waves in the baryon-like state. The quark distributions depicted in {\bf(a)} and {\bf(b)} represent the configurations under the strong and weak correlation assumptions, respectively. In the strongly correlated configuration {\bf(a)}, the confined quarks have almost the same size of momenta so that the distribution of $u$ quarks is concentrated around $k_{F}^{d}$. However, in the weakly correlated configuration {\bf(b)}, $k_{\textrm{conf.}}^{u}$ is broadly distributed as the confined quark momenta can be deviated from each other.}
\label{fig1}
\end{figure}

The isospin asymmetry appears  naturally under consideration of electroweak interactions and subsequent equilibrium conditions. This asymmetric configuration can arise in either the baryons and quarks. The details will be strongly dependent on $k^b_F$ as the physical constraints between the baryons and quarks are related through the shell-like baryon distribution.  $k^b_F$ can be imagined differently depending on the assumption of the confined quark state distributed closely to the saturated quark Fermi surface. As illustrated in Fig.~\ref{fig1}(a), $k^b_F$ should show weak dependence on the flavor asymmetry if the confined quark momenta around the saturated quark Fermi surface are strongly correlated, while it can depend strongly on the asymmetry if the confined quark momenta are weakly correlated [Fig.~\ref{fig1}(b)]. We propose two phenomenological approaches by assuming the quark momentum correlation strength $r^{s/w}_{q_1 q_2}$ and the weight function $\omega_{s/w} \left(x\right)$, where $s$ and $w$ denote the strong and weak correlations respectively. As an example, one may imagine an isospin asymmetric configuration where  $q_1$ quarks are saturated first as total baryon density increases and $q_2$ quarks follow after ($q_1 = d,~ q_2=u$ in Fig.~\ref{fig1}). Then, the momentum of $q_2$ quark confined in the lower boundary of the baryon shell can be supposed as follows:
\begin{align}
k_{\textrm{conf.}}^{q_2} &= k_F^{q_1} + r^{s/w}_{q_1 q_2} w_{s/w} \left( k_F^{q_2} - k_F^{q_1} \right).
\end{align} 
$r^{s/w}_{q_1 q_2}$ determines the correlation strength and the correlation weight function $w_{s/w}(x)$ satisfies the boundary condition [$w_{s/w}(x \ll \Lambda_{\textrm{QCD}}) = 0$, $w_{s/w}(x \gg \Lambda_{\textrm{QCD}}) \rightarrow 1$]. The lower boundary of the shell-like distribution can be written as
\begin{align}
k_F^n =~&k_{\textrm{conf.}}^{u} + 2 k_{\textrm{conf.}}^{d} \nonumber \\ 
=~& \Theta (k_F^d - k_F^u) \left( 3k_F^d + r^{s/w}_{qq} w_{s/w}\left( k_F^d - k_F^u \right) \right) + \Theta (k_F^u - k_F^d) \left( 3k_F^u+2r^{s/w}_{qq}w_{s/w}\left( k_F^u - k_F^d \right) \right),\\
k_F^p  =~& 2 k_{\textrm{conf.}}^{u} + k_{\textrm{conf.}}^{d}\nonumber  \\
=~& \Theta (k_F^d - k_F^u) \left( 3k_F^d + 2 r^{s/w}_{qq} w_{s/w}\left( k_F^d - k_F^u \right) \right)  + \Theta (k_F^u - k_F^d) \left( 3k_F^u+  r^{s/w}_{qq} w_{s/w} \left( k_F^u - k_F^d \right) \right),\\
k_F^{\Lambda} =~&k_{\textrm{conf.}}^{u} +  k_{\textrm{conf.}}^{d}  +  k_{\textrm{conf.}}^{s} \nonumber \\
=~&\Theta (k_F^d - k_F^s) \left( 3 k_F^d+ r^{s/w}_{qq} w_{s/w}\left( k_F^d - k_F^u \right) + r^{s/w}_{qs} w_{s/w}\left( k_F^d - k_F^s \right)\right)\nonumber\\
&~+ \Theta (k_F^s - k_F^d)\left( 3k_F^s+ r^{s/w}_{qs} w_{s/w}\left( k_F^s - k_F^d \right) +  r^{s/w}_{qs} w_{s/w}\left( k_F^s - k_F^u \right) \right)\label{sb3},
\end{align}
where $r^{s/w}_{qq}$ and $r^{s/w}_{qs}$ determines the momenta correlation strength between the up-down quarks and light-strange quarks respectively. For both the strongly and weakly correlated assumptions, we assume the momenta of light quarks are even more strongly correlated than in the light-strange quarks case ($\vert r^{s/w}_{qq}\vert < \vert r^{s/w}_{qs} \vert $) by following the phenomenological understanding of the EMC effect~\cite{Aubert:1983xm, Hen:2016kwk}.  In Eq.~\eqref{sb3}, $k_F^d \geq k_F^u$, $k_F^s \geq k_F^u$ conditions are understood from the electromagnetic charge and the weak decay channel of  the quarks. Detailed descriptions for the two assumptions follow.

\subsubsection{Assumption I: Strongly correlated momentum of confined quark}\label{sec2b1}

\begin{figure}
\includegraphics[width=0.88\textwidth]{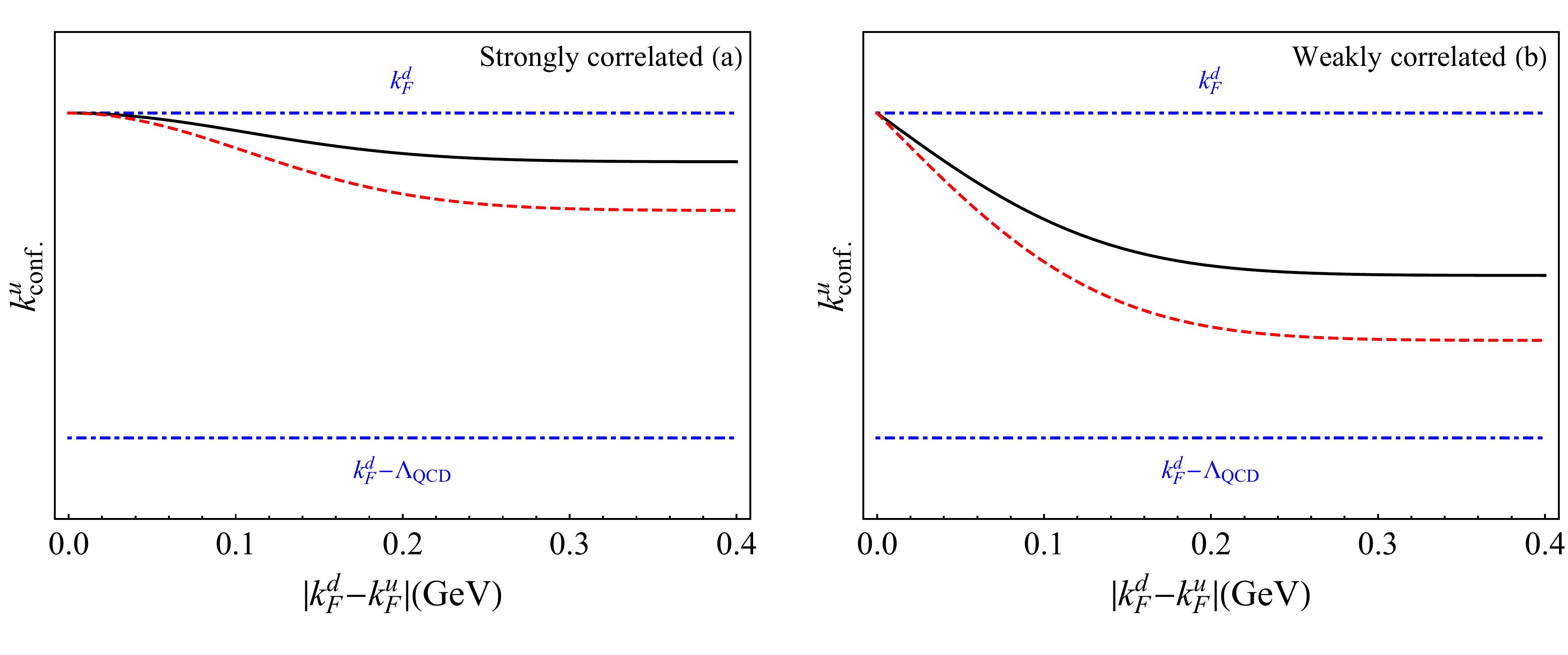} 
\caption{Illustration of $k_{\textrm{conf.}}^u$ with $k_F^d>k_F^u$ condition. {\bf{Left} (a)}: $k_{\textrm{conf.}}^u$ under the strong correlation assumption is plotted with the black solid (red dashed) line and $r^{s}_{qq}=-30~\textrm{MeV}$ ($r^{s}_{qq}=-60~\textrm{MeV}$). The confined quark momenta are closely located around $k_F^d$ and are weakly dependent on $\vert k_{F}^d - k_{F}^u  \vert$. {\bf{Right} (b)}: $k_{\textrm{conf.}}^u$ under the weak correlation assumption is plotted with the black solid (red dashed) line and $r^{w}_{qq}=-100~\textrm{MeV}$ ($r^{w}_{qq}=-140~\textrm{MeV}$). The confined quark momenta rapidly deviate away from $k_F^d$ as $\vert k_{F}^d - k_{F}^u  \vert$ becomes large.} 
\label{fig2}
\end{figure}

In the large-$N_c$ limit, one may assume a strong correlation between the momenta of confined quarks because the confinement mechanism should be very similar to the one of the hadron state in vacuum. This clear confinement should occur even for the quarks whose momentum is distributed closely around the saturated Fermi surface, where the occupation number is almost 1. In a simplest guess for the constituent quarks of a baryon, one can imagine the confined quarks sharing a same typical momentum $k_{\textrm{conf.}}^Q=k^{b}/3$ balanced by the internal interaction. If all the quark momenta are strongly correlated as in this simple guess, the difference between the momenta of two confined quarks should be minimal, even though the flavor asymmetry of saturated quarks becomes large ($\vert k_{\textrm{conf.}}^{q_i} - k_{\textrm{conf.}}^{q_j}  \vert \ll \Lambda_{\textrm{QCD}},~k_{\textrm{conf.}}^{q_i} > k_{F}^{q_i}$, and $ \vert k_{F}^{q_i} - k_{F}^{q_j}\vert >\Lambda_{\textrm{QCD}}$,  where $i,j$ denote the quark flavors). To depict this nature, one can suppose a slowly varying weight function: 
\begin{align}
w_s(x)& =1- \exp{ \left(-\vert x \vert^2/{\delta}^2 \right)},\label{cor1}
\end{align}
where ${\delta}=0.15~\textrm{GeV}$ determines the non-trivial range where $w_s(x<\Lambda_{\textrm{QCD}}) < 1$. One can find the required flavor asymmetry insensitivity at $x= \vert k_{F}^{q_i} - k_{F}^{q_j}\vert < \Lambda_{\textrm{QCD}}$ in the derivative of Eq.~\eqref{cor1}:
\begin{align}
\frac{d w^{}_s(x)}{dx}& = \frac{ 2 x }{\delta^2} \exp{ \left(-\vert x \vert^2/{\delta}^2 \right)},\label{dcor1} 
\end{align}
where the factor $  2 x /{\delta}^2 $ suppresses the derivative at small $x$. If one assigns a small negative  $r^{s}_{q_1 q_2}$, the minimal difference $ \vert k_{\textrm{conf.}}^{q_2} - k_{\textrm{conf.}}^{q_1}  \vert \ll \Lambda_{\textrm{QCD}}$ will be guaranteed even at  $ \vert k_{F}^{q_i} - k_{F}^{q_j}\vert > \Lambda_{\textrm{QCD}}$. As illustrated in Fig.~\ref{fig2}(a), $k_{\textrm{conf.}}^u \rightarrow k_F^d + r^{s}_{qq}$ in the $\vert k_{F}^d - k_{F}^u  \vert > \Lambda_{\textrm{QCD}}$ limit and $k_{\textrm{conf.}}^u$ varies slowly even with  a relatively large $\vert r^{s}_{qq} \vert$, which implies $k_{F}^{n,p} \simeq N_c \textrm{max.} \left[ k_{F}^{u}, k_{F}^{d}\right]$. Under this assumption, one can anticipate the minimal flavor asymmetry in the quark Fermi sea: populating a specific flavor of quark leads to the large shift of the shell-like distribution $k_F^b \simeq N_c \textrm{max.} \big[ k_F^{Q_i}\big] $.

\subsubsection{Assumption II: Weakly correlated momentum of confined quark}\label{sec2b2}

On the other hand, one can imagine the weakly correlated momenta of the confined quarks in the confinement range $ \vert k^{q_i}_{\textrm{conf.}} - k^{q_j}_{\textrm{conf.}}  \vert \lesssim \Lambda_{\textrm{QCD}}$. If one considers the nonzero chiral condensate in the confined baryon phase and the symmetry restoration at high density regime~\cite{Cohen:1991nk, Hatsuda:1992bg, Brown:2001nh, Kaiser:2007nv, Kaiser:2008qu,  Kojo:2009ha, Fukushima:2010bq}, the confinement mechanism of the quarks distributed closely around the saturated Fermi surface would be quite different from the one of the vacuum case where the symmetry is broken. The confined state would look like the correlated state of three non-perturbative quarks whose ground energy scale is $m_b\simeq 1~\textrm{GeV}$, rather than the clearly distinguishable baryon state. In this weakly correlated assumption, if the confined quark momentum of a specific flavor becomes enhanced by saturation ($k_{\textrm{conf.}}^{q} > k_{F}^{q}$) so that the flavor asymmetry becomes large, the other confined quarks can take some lower unoccupied phase space ($k_{F}^{q}> k_{\textrm{conf.}}^{q_i}> k_{F}^{q}-\Lambda_{\textrm{QCD}}$) to minimize the ground state energy [Fig.~\ref{fig1}(b)]. For this weakly correlated assumption, the following weight function can be supposed:
\begin{align}
w_w(x)&=\textrm{erf}{\left(-\vert x \vert/{\delta} \right)},\label{cor2}
\end{align}
where $\textrm{erf}(x)$ denotes the error function and ${\delta}=0.15~\textrm{GeV}$ has the same role given in Eq.~\eqref{cor1}. As one can find in the following derivative, this weight function converges rapidly to 1:
\begin{align}
\frac{d w^{}_w(x)}{dx} &= \frac{2 } {\sqrt{\pi}\delta}\exp{ \left(-\vert x \vert^2/\delta^2 \right)}\label{dcor2}.
\end{align}
Because it does not have the factor $x$  in comparison with Eq.~\eqref{dcor1}, the weight function~\eqref{cor2} is much more sensitive to $x= \vert k_{F}^{q_i} - k_{F}^{q_j}\vert$ even in the small $x$ regime . With a large negative $r^{w}_{q_1 q_2}$, the non-negligible difference $ \vert k_{\textrm{conf.}}^{q_2} - k_{\textrm{conf.}}^{q_2}  \vert < \Lambda_{\textrm{QCD}}$ can be obtained. As can be found in Fig.~\ref{fig2}(b), the error function enables relatively fast reduction of $k_{\textrm{conf.}}^u$ even at small $\vert k_{F}^d - k_{F}^u  \vert$. Compared to the case of the strong correlation assumption, relatively larger flavor asymmetry is anticipated among the saturated quarks: if $k_{\textrm{conf.}}^{q_2}$ is distributed away from $k_F^{q_1}$, $k_F^{b}$ can have $k_F^b < N_c \textrm{max.} \big[ k_F^{Q_i}\big] $ for large $\vert k_{F}^{q_1} - k_{F}^{q_2}  \vert$.

\section{Equation of state and application to Neutron stars}\label{sec3}

{\bf Equilibrium constraints and parameter set.} In the three-flavor system with electron clouds, the physical configuration should be constrained by the baryon number conservation, charge neutrality, and possible weak interactions. As summarized in Ref.~\cite{Duarte:2020xsp},  the weak interactions leads to following constraints:
\begin{align}
\mu_n&=\mu_p + \mu_e,\label{beq1}\\
\mu_{\tilde{d}}&=\mu_{\tilde{u}}+N_c \mu_e,\label{beq2}\\
\mu_n&=\mu_{\Lambda}~(\textrm{when}~n_{\Lambda}\neq0.~ n_{\Lambda}=0~\textrm{if}~\mu_{\Lambda}<m_{\Lambda}),\label{beq3}\\
\mu_{\tilde{d}}&=\mu_{\tilde{s}}~(\textrm{when}~n_{\tilde{s}}\neq0.~n_{\tilde{s}}=0~\textrm{if}~\mu_{\tilde{s}}<N_c m_{s}),\label{beq4}
\end{align}
where $\mu_{\tilde{Q}_i}=N_c \mu_{Q_i}$ denotes the quark chemical potential in units of baryon number.  The saturated quarks on the Fermi surface are allowed to decay onto the other Fermi surface of different flavor. Under baryon number conservation and charge neutrality, these constraints lead to the dynamical equilibrium condition: 
\begin{align}
&\textrm{if}~n_{\Lambda}=0,~n_{s}=0,~ \mu_n= N_c \mu_{d} -\mu_e =\mu_p + \mu_e,\label{deq1}\\
&\textrm{if}~n_{\Lambda}\neq0,~n_{s}\neq0,~ \mu_n= N_c \mu_{d} -\mu_e = \mu_{\Lambda}=\mu_p + \mu_e= N_c \mu_{s} -\mu_e,\label{deq2}\\
&\textrm{if}~n_{\Lambda} = 0,~n_{s}\neq0,~ \mu_n= N_c \mu_{d} -\mu_e = \mu_{\Lambda}=\mu_p + \mu_e,\label{deq3}\\
&\textrm{if}~n_{\Lambda}\neq0,~n_{s}=0,~ \mu_n= N_c \mu_{d} -\mu_e =\mu_p + \mu_e= N_c \mu_{s} -\mu_e,\label{deq4}
\end{align}
where $\mu_n= N_c \mu_{d} -\mu_e $ is the generalization of the dynamical equilibrium constraint $\mu_N = N_c \mu_q$ in the isospin symmetric configuration~\cite{Jeong:2019lhv}.
Hereafter, we will calculate all the physical quantities under the constraints.  The following numbers will be used as the representative parameter set: $N_c=3$ for the number of colors, \{$m_{n,p}=1~\textrm{GeV}$,  $m_{\Lambda}=1.2~\textrm{GeV}$, $m_{q}=0.333~\textrm{GeV}$, and $m_{s}=0.533~\textrm{GeV}$\} for the fermion masses\footnote{The mass number set is chosen to follow the previous work~\cite{Duarte:2020xsp}. The difference from the calculation with the physical mass set \{$m_{n,p}=0.938~\textrm{GeV}$,  $m_{\Lambda}=1.115~\textrm{GeV}$, $m_{q}=0.313~\textrm{GeV}$, and $m_{s}=0.490~\textrm{GeV}$\} is negligible.}, \{$r^{s}_{qq}=-30~\textrm{MeV}$ and $r^{s}_{qs}=-60~\textrm{MeV}$\} for the strong correlation assumption, \{$r^{w}_{qq}=-100~\textrm{MeV}$ and $r^{w}_{qs}=-140~\textrm{MeV}$\} for the weak correlation assumption, $n_0 = 6 \rho_0 $ for the hard-core density ($r_c\simeq 0.6~\textrm{fm}$), and corresponding regulator $\Lambda_{Q}=180~\textrm{MeV}$ which attenuates the unphysical noise.

\subsection{Density profile of particles in the excluded volume model with shell-like baryon distribution}\label{subs3a}

\begin{figure}
\includegraphics[width=0.92\textwidth]{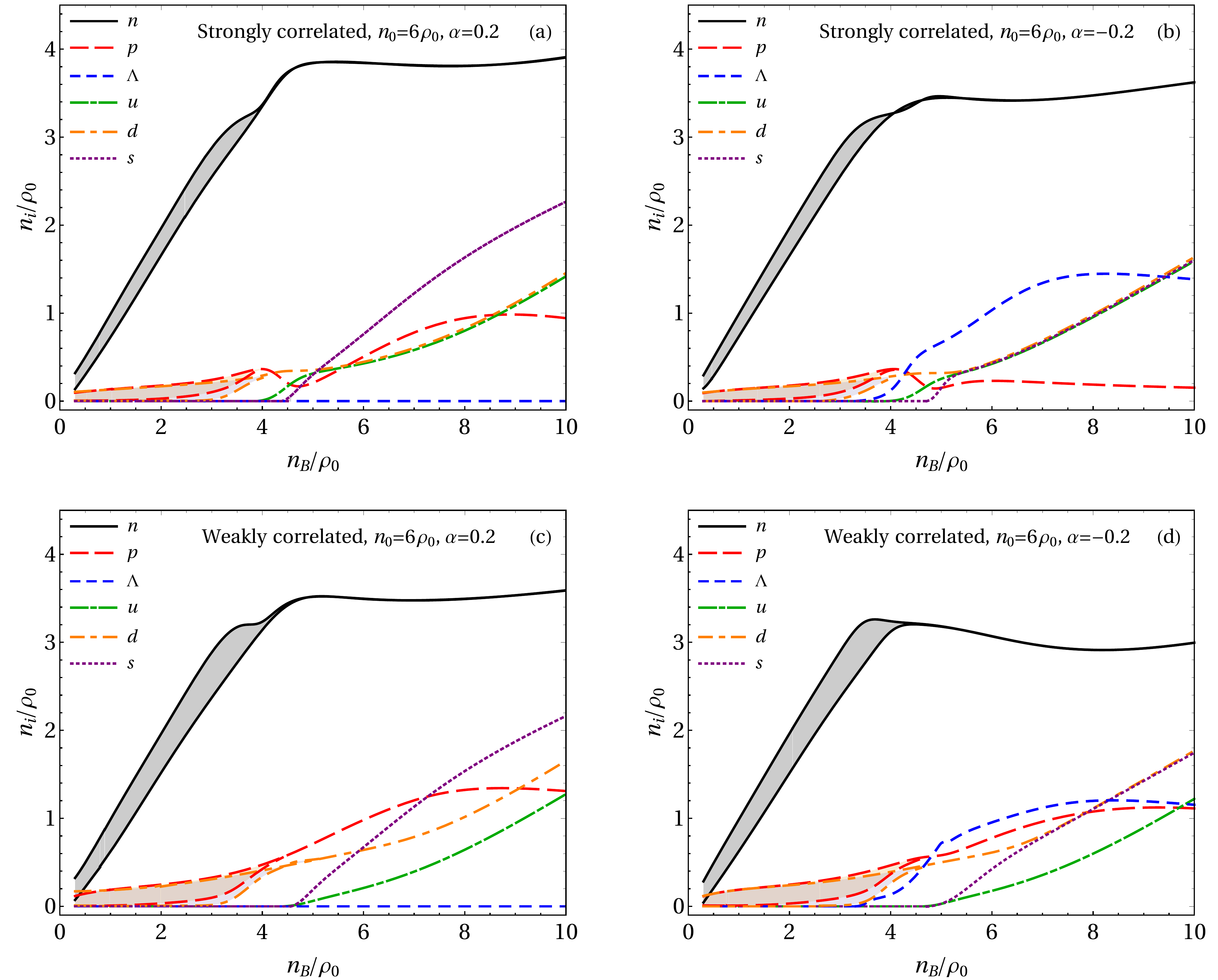} 
\caption{Density profiles of fermions for $n_0 = 6\rho_0$, and $\Lambda_{Q}=0.18~\textrm{GeV}$.  The profiles in the upper {\bf (a, b)} and lower {\bf(c, d)} sides are obtained under the {\bf strong} and {\bf weak} correlation assumptions, respectively. The profiles in the left {\bf(a, c)} and right {\bf(b, d)} sides are obtained under $\alpha=0.2$ and  $ \alpha=-0.2 $ conditions, respectively. The shaded band denotes the possible deviation depending on the saturation moment of the $d$ quark Fermi sea, which may rely on the proper EoS which covers the low density regime.} 
\label{fig3}
\end{figure}

We present the density profile of particles to understand the complicated dynamical properties from the shell-like distribution. The density profiles of particles are plotted in Fig.~\ref{fig3} under $n_0 = 6\rho_0$, and $\Lambda_{Q}=180~\textrm{MeV}$ conditions. As one can find in the profiles (a) and (c) of Fig.~\ref{fig3}, the stronger repulsive core ($\alpha = 0.2$) for the $\Lambda$ hyperon suppresses the emergence of the $\Lambda$ degree of freedom even in the high density regime while the weaker repulsive core ($\alpha = -0.2$) allows $n_{\Lambda}>0$ in the same regime as shown in the profiles (b) and (d). One can find the reason from the $\omega_{\Lambda}$ dependent terms of the baryon chemical potential~\eqref{bchmp}: it becomes hard to satisfy the equilibrium constraint~\eqref{beq3} with the other constraints in a simultaneous way because $\mu_{\Lambda}$ is enhanced by $\omega_{\Lambda}>1$. 

Meanwhile, when the $\Lambda$ degree of freedom is suppressed, the $s$ quark takes relatively larger portion than in the cases where $n_{\Lambda}>0$ with $\alpha = -0.2$ [(b) and (d) of Fig.~\ref{fig3}]. As can be found in Eq.~\eqref{qchemp}, the quark chemical potential has a contribution from the shell-like distribution if the quark Fermi momentum is related to the confined quark momentum  via the Pauli exclusion principle. Because there is no $\Lambda$ shell-like distribution in the $\mu_{\tilde{s}}$ for $\alpha = 0.2$ case, relatively large $n_{\tilde{s}}$ can be accommodated, satisfying the constraint~\eqref{beq4} where $\mu_{\tilde{d}}$ has the contribution from the $n,p$ shell. By the same reason, one can understand the difference between the profiles from the strong and weak correlation assumptions. Under the strong correlation of the confined quark momenta, the large isospin asymmetry in the quark Fermi sea enhances the lower boundary of nucleon momentum as $k^{n,p}_F \simeq 3 k_F^d$, by which the nucleons in the shell obtain huge energy enhancement. Thus, it is dynamically favored for the isospin symmetric configuration of the light quark Fermi sea by the constraint~\eqref{deq3} [Fig.~\ref{fig3}(a)]. If the $\Lambda$ shell ($n_{\Lambda}>0$) exists, all the quark Fermi sea becomes almost symmetric and the asymmetric configuration only appears in the shell-like distribution of the baryon side by the constraint~\eqref{deq2} in the high density regime [Fig.~\ref{fig3}(b)]. However, if the confined quark momenta are weakly correlated, a large favor asymmetry is allowed for the quark Fermi sea in the same density regime [see the profiles (c) and (d) in Fig.~\ref{fig3}] as $k^{n,p}_F \leq 3 k_F^d$ and $k^{\Lambda}_F \leq 3 k_F^s$. Therefore, the constraints~\eqref{deq2} and \eqref{deq3} can be satisfied in that asymmetric configuration.

In all the cases, the quark Fermi sea is saturated in order of $d$, $u$, and $s$ quark flavors. After the saturation, ($n_B \geq 5\rho_0$), each baryon density profile appears to converge to the asymptotic number and the quark Fermi sea takes all the increment of the baryon number density ($d  n_{B} \simeq d n_{\tilde{Q}}$). By definition, this model does not contain the essential attractive and repulsive potentials required to reproduce the low density properties of nuclear matter. The saturation moment of the $d$ quark Fermi sea can differ, by the proper modifications, to acquire the low density properties. The expected possible configurations in the low density regime are denoted as the shaded area in Fig.~\ref{fig3}. The qualitative behavior of the density profile does not change when a different hard-core density $n_0 = 5\rho_0$ is assigned.

\subsection{Equation of state and speed of sound}\label{subs3b}

\begin{figure}
\includegraphics[width=0.95\textwidth]{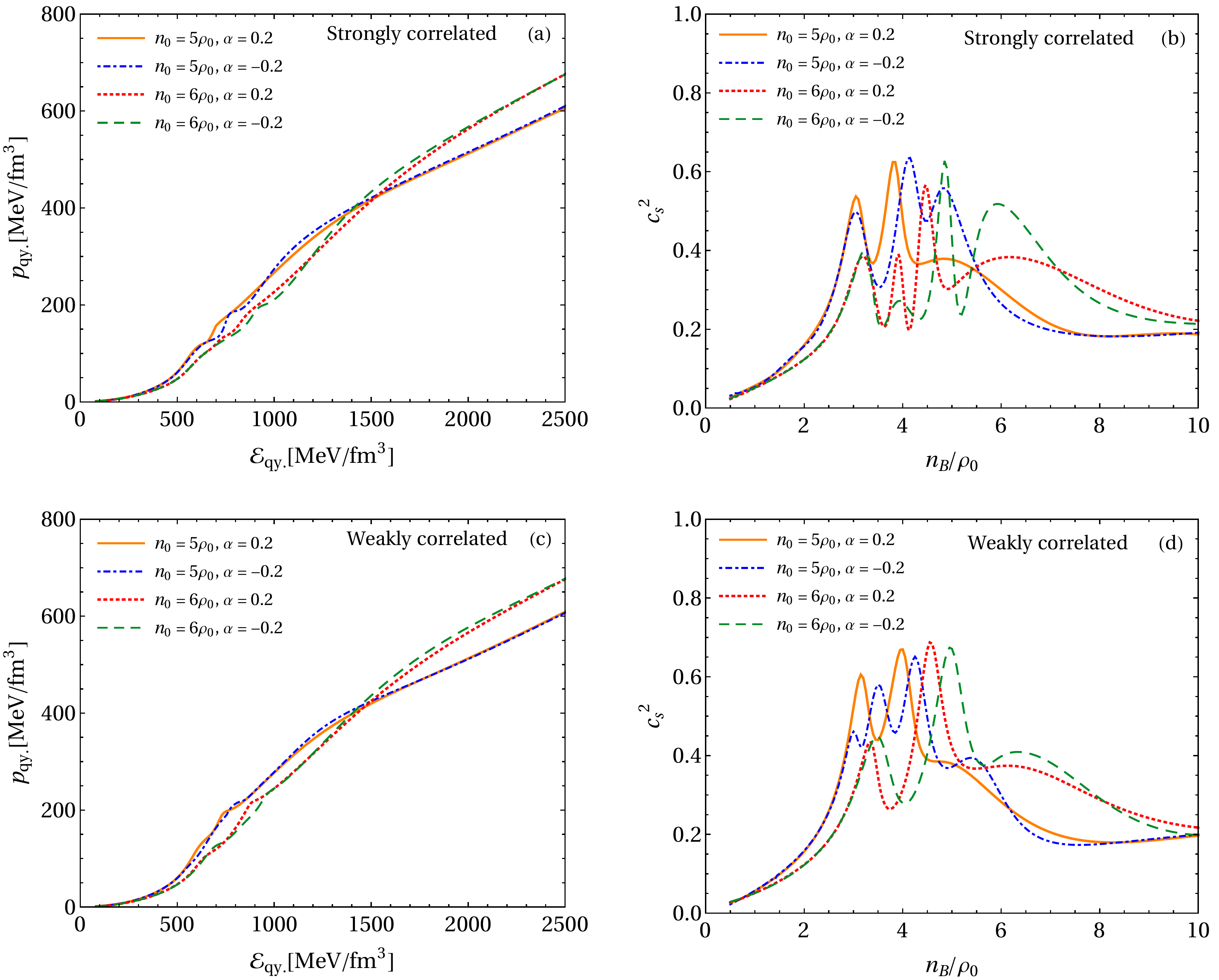} 
\caption{Evolution of EoS {\bf(left)} and corresponding sound velocity {\bf(right)}. The plots in the upper {\bf(a, b)} and lower {\bf(c, d)} sides are obtained under the {\bf strong} and {\bf weak} correlation assumption, respectively. The stiffly increasing segments in {\bf(a, c)} correspond to the onset moments of the new degrees of freedom and subsequently expanding shell-like phase structure of the baryons. The peaks in {\bf(b, d)} corresponds to the stiff behavior of the EoS in {\bf(a, c)}.} 
\label{fig4}
\end{figure}

In the zero-temperature limit, the pressure and corresponding sound velocity can be found as
\begin{align}
p_{\textrm{qy.}}&=- \varepsilon_{\textrm{qy.}} +\mu_{B} n_{B},\\
c_s^2&=\frac{\partial p_{\textrm{qy.}}}{\partial \varepsilon_{\textrm{qy.}} }= \frac{n_{B}}{\mu_{B} \frac{\partial n_B}{\partial \mu_B}}.
\end{align} 
As can be found in the Figs.~\ref{fig4}(a) and ~\ref{fig4}(b), the EoS increases stiffly around the saturation moment of the quark Fermi sea. The evolution becomes more stiff when the stronger repulsive core ($n_0 = 5\rho_0$) is considered. Because each quark flavor can be saturated separately  in this system, there are several stiffly increasing segments in the evolution curve of the EoS. This tendency does not appear when all the quark flavors simultaneously saturate~\cite{Jeong:2019lhv, Zhao:2020dvu}. This evolution looks similar to the results presented in Refs.~\cite{Marczenko:2020jma, Marczenko:2020wlc} where the first-order phase transition is implied via the hybrid quark-meson-nucleon model~\cite{Benic:2015pia, Marczenko:2017huu, Marczenko:2020jma, Marczenko:2020wlc}. However, as one can find in the sound velocity plots [Figs.~\ref{fig4}(b) and (d)], $c^2_s>0$ in the stiffly increasing segment, which means that our model does not present the first-order phase transition around the onset of quark sea saturation. In the high density regime, the stiffness becomes moderated and appears to converge to the relativistic ideal limit ($c^2_s =1/3$) as one can anticipate from the definition of the model~\eqref{defed}. Although the stiffness converges to the ideal limit, the weaker repulsive core ($n_0 = 6\rho_0$) leads to a harder EoS in the high density regime because the system can accommodate more baryons under the weaker repulsion. 

The details of the stiff increments of EoS can be understood from the corresponding sound velocity plots and the density profiles. Under the strong correlation assumption for the confined quark momenta, one can read the overall stiffness of EoS from the the peak value of sound velocity as $\textrm{max.}[c^2_s] \simeq 0.6$, while its peak becomes $\textrm{max.}[c^2_s] \simeq 0.7$ under the weak correlation assumption. The multiple peaks and the scale of sound velocity are compatible with the results from Refs.~\cite{Marczenko:2020jma, Marczenko:2020wlc, Motornenko:2019arp}. As discussed in Sec.~\ref{subs3a}, the flavor asymmetry of the quark Fermi sea evidently appears under the weak correlation assumption, which can make relatively large energy enhancement to the baryon shell side. In comparison of the peaks in the sound velocity, the scale of the final bump after the $s$ quark saturation is determined by the existence of the $\Lambda$ shell-like distribution. After the saturation of all the quark Fermi sea, the shell-like distributions of the baryons  expand rapidly as the saturated quarks take almost all of the total baryon number density increment $\partial n_{B}/\partial n_{\tilde{Q}} \simeq 1$, so that $k^b_F\simeq N_c k^Q_F$. If $n_{\Lambda}>0$ ($\alpha = - 0.2$), the expanding $\Lambda$ shell makes the bump slightly larger than the case of no $\Lambda$ shell ($\alpha =  0.2$). The locations of early appearing peak can be altered by the phenomenological modification of the EoS for the low density regime as the saturation moment of the $d$ quark Fermi sea may depend on the modification.

\subsection{Mass-radius relation of quarkyonic neutron star}

\begin{figure}
\includegraphics[width=0.95\textwidth]{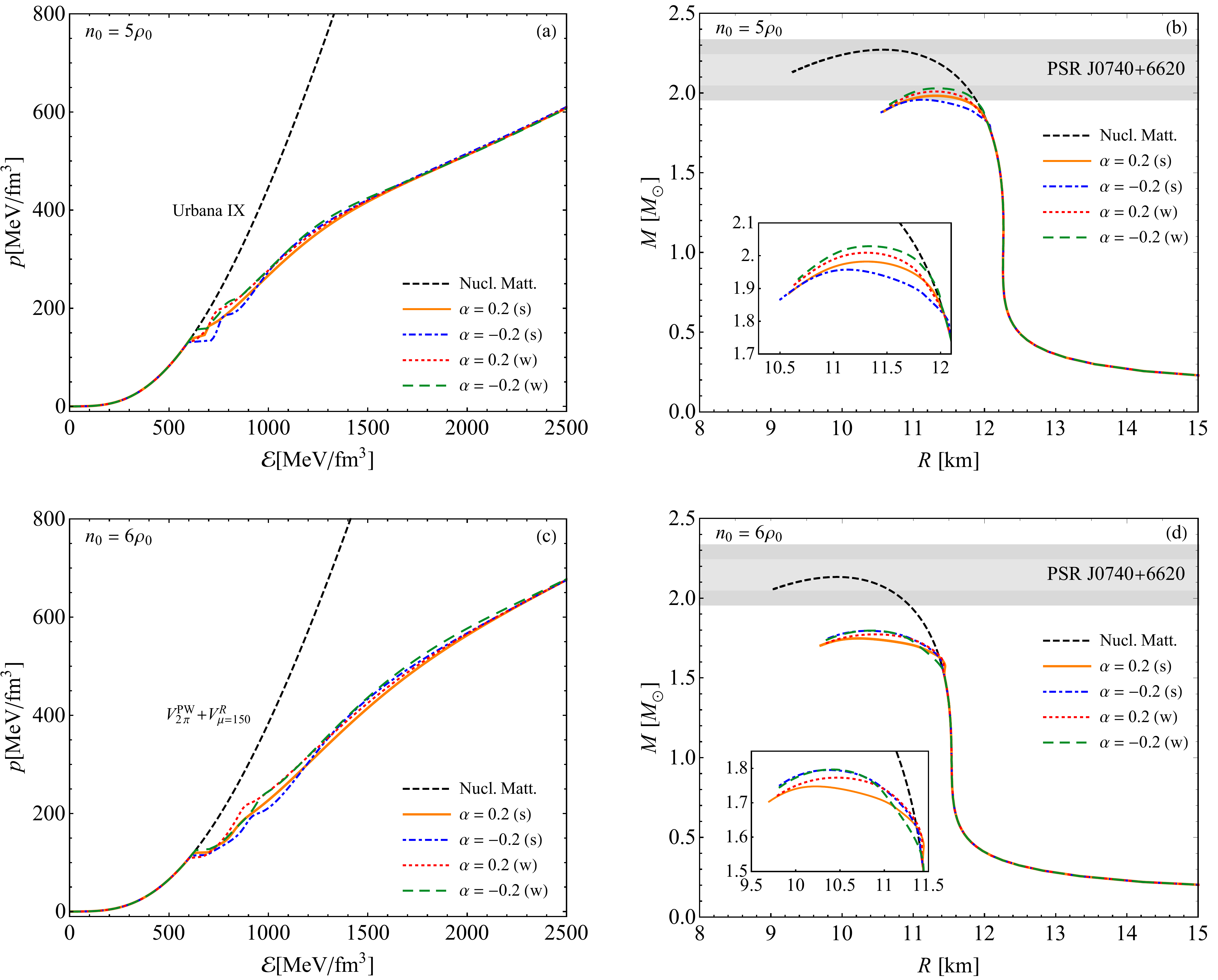} 
\caption{Maxwell constructions of EoS {\bf(left)} and mass-radius relations from the solution of TOV equations {\bf(right)}. Abbreviations {\bf (s)} and {\bf (w)} represent the strong and weak correlation assumptions for the confined quark momenta, respectively. The crust EoS~\cite{Baym:1971pw, Negele:1971vb} is used for the lower density regime of $ 0\leq \rho_B \leq 0.5 \rho_0$. In the density range of $ 0.5 \rho_0 \leq \rho_B \leq \rho_M$, two different parameter sets~\cite{Gandolfi:2013baa} are used for the nuclear EoS: \{Urbana IX force: $\tilde{a}=-28.3~\textrm{MeV},~\tilde{b}=10.7~\textrm{MeV}$\} for the plots in {\bf (a, b)} and \{$V^{PW}_{2\pi}+V^{R}_{\mu=150}$: $\tilde{a}=-29.8~\textrm{MeV},~\tilde{b}=13.6~\textrm{MeV}$\}  for the plots in {\bf (c, d)}. The EoS of the excluded-volume model is used for the higher density regime beyond $\rho_M$.  Stiffer evolution with $n_0=5\rho_0$ {\bf (a, b)}: $M_{\textrm{max.}}=2.03 M_{\odot}$ and $R_{1.4}=12.5~\textrm{km}$. Softer evolution with $n_0=6\rho_0$ {\bf (c, d)}: $M_{\textrm{max.}}=1.8 M_{\odot}$ and $R_{1.4} = 11.5~\textrm{km}$. The gray shaded inner (outer) band represents the 68.3\% (95.4\%) credence  range of $M_{\textrm{max.}}$ estimated from Ref.~\cite{Cromartie:2019kug}.} 
\label{fig5}
\end{figure}

Now we can explore the possible quarkyonic configuration in the compact stellar state by solving the Tolman-Oppenheimer-Volkof (TOV) equations~\cite{Tolman:1939jz, Oppenheimer:1939ne}:
\begin{align}
\frac{d p(r)}{dr} &= \frac{ G[\epsilon(r)+p(r)  ] [M(r)+4\pi r^3 p(r) ]}{r[r-2GM(r)]},\\
\frac{d M(r)}{dr} & = 4\pi r^2 \epsilon(r),
\end{align} 
where $G$ is the gravitational constant and the boundary conditions $p(R_{\textrm{star}})=0$ and $M(R_{\textrm{star}})=M_{\textrm{star}}$ are assumed. To estimate physically reasonable mass-radius relation, the low density part of our model needs correction as it does not contain the essential attractive and repulsive contributions required to describe the low density nuclear matter properties. The EoS of our model will be kept from the intermediate regime to the high density limit because we assume that the hard-core repulsive interaction will dominate the other contributions in the intermediate density regime, and we focus on the role of the dynamically generated shell-like distribution of baryons in the high density regime.  Below a critical density (say $n_B \leq \rho_M$), some proper EoS can be adopted instead of introducing additional mean-field potentials. In the extremely low density regime ($ 0\leq n_B \leq 0.5 \rho_0$), the EoS of the outer crust~\cite{Baym:1971pw, Negele:1971vb} will be used. Since the low density configuration of our model can be simply regarded as pure neutron matter (see the profiles in Fig.~\ref{fig3}), the nuclear EoS developed for neutron rich matter~\cite{Gandolfi:2013baa} can be quoted for the intermediate density regime ($0.5\rho_0 \leq n_B \leq \rho_M$) as
\begin{align}
E/A= \left({p^{n}_{F}}^2+m_n^2 \right)^{\frac{1}{2}}-m_n+\tilde{a}\left( \frac{n_n}{\rho_0}\right) + \tilde{b}\left(\frac{n_n}{\rho_0} \right)^2,
\end{align} 
where $p^{n}_{F}=(3\pi^2 n_n)^{1/3}$ denotes the neutron Fermi momentum in the ideal gas limit. The attractive ($\tilde{a}$) and repulsive ($ \tilde{b}$) coefficients have been determined by the possible three-body nucleon forces. The parameter sets \{Urbana IX force: $\tilde{a}=-28.3~\textrm{MeV},~\tilde{b}=10.7~\textrm{MeV}$\}  and \{$V^{PW}_{2\pi}+V^{R}_{\mu=150}$: $\tilde{a}=-29.8~\textrm{MeV},~\tilde{b}=13.6~\textrm{MeV}$\} are used in the stiffer and softer nuclear EoS, respectively. The stiffer (softer) nuclear EoS is interpolated with our high density EoS with $n_0=5\rho_0$ ($n_0=6\rho_0$), requiring a minimal  Maxwell construction interval [$P_{\textrm{nucl.}}(\mu_M) = P_{\textrm{qy.}}(\mu_M)$ where $\mu_M = \mu_B(\rho_M)$].  Comparing the Maxwell construction plotted in Figs.~\ref{fig5}(a) and \ref{fig5}(b) to the EoSs plotted in Fig~\ref{fig4}, one can find that the procedures are done around $3\rho_0 \leq \rho_M \leq 3.5\rho_0$. Since the $\Lambda$ hyperons and $s$ quarks are generated beyond $ \rho_M$, the interpolated EoS conserves the particle density profiles of the original quarkyonic system. For the interpolated EoS obtained under the $n_0=5\rho_0$ condition [Figs.~\ref{fig5}(a, b)], the moment of interpolation differs due to the assumptions regarding the high density EoS. Among the interpolated curves plotted in Fig.~\ref{fig5}(a), the $\alpha=0.2$ cases appears to be interpolated smoothly and the interval which looks like first-order transition appears minimally, for while the $\alpha=-0.2$ cases show a non-negligible interval. Although the saturation moments of the light quark Fermi sea are included in the non-negligible interval with the $\alpha=-0.2$ cases, one can still regard the interpolated EoS as the effective quarkyonic-like model because the saturated quark number density is quite small indeed in the interval. These Maxwell construction intervals are the artifacts of the low density EoS estimation procedure.\footnote{Two possible methods for improvement are briefly suggested in the following section.}

Corresponding mass radius relations are presented in Fig.~\ref{fig5}(b). The low mass stage is governed by the nuclear EoS from the Urbana IX model and the high mass tails are determined by the quarkyonic-like excluded-volume model. The highest mass state appears as \{$M_{\textrm{max.}}=2.03 M_{\odot}$, $ R_{M_{\textrm{max.}}}=11.4~\textrm{km}$\} where the weaker repulsive core of $\Lambda$ ($\alpha=-0.2$) and the weakly correlated confined quark momenta are assumed. The other curves are barely located in the possible range estimated from the recent observation~\cite{Cromartie:2019kug} ($M_{\textrm{max.}}=2.14^{+0.20}_{-0.18} M_{\odot}$). The constraint from the GW observations, $R_{1.4}=12.5~\textrm{km} < 13.5~\textrm{km}$~\cite{TheLIGOScientific:2017qsa, Abbott:2018exr}, is satisfied by the adopted nuclaer EoS and $R_{1.8}=12.2~\textrm{km} < 15~\textrm{km}$~\cite{Abbott:2020uma} is satisfied by the excluded-volume model.
 
If the $n_0=6\rho_0$ condition is considered [Figs.~\ref{fig5}(c) and \ref{fig5}(d)], the interpolation interval appears minimally in all the cases and the Maxwell construction is done before the saturation of the $d$ quark Fermi sea. The low mass stage is governed by the nuclear EoS with the $V^{PW}_{2\pi}+V^{R}_{\mu=150}$ potential. While the interpolated EoS can be regarded as the quarkyonic-like model, $M_{\textrm{max.}}=2.14^{+0.20}_{-0.18} M_{\odot}$ cannot be reproduced from the EoS. The highest mass state appears as \{$M_{\textrm{max.}}=1.8 M_{\odot}$, $ R_{M_{\textrm{max.}}} = 10.4~\textrm{km}$\} in both cases of the weaker repulsive core of $\Lambda$ ($\alpha=-0.2$). The other curves present the maximal mass around   $M_{\textrm{max.}}\simeq 1.75 M_{\odot}$ and the corresponding radius in the range of $10~\textrm{km} \leq R_{M_{\textrm{max.}}} \leq 10.5~\textrm{km}$.

In comparison with the previously reported work~\cite{Duarte:2020xsp}, one can find that the formation of the shell-like distribution of the baryon state by the Pauli exclusion principle can make the EoS hard enough to support the large mass state of a neutron star. Although it is necessary to adopt the nuclear EoS for the lower density regime, the higher mass state evolution is determined by the EoS of the quarkyonic-like excluded-volume model.   In the curves plotted in Fig.~\ref{fig5}(b), the deviation point of the higher mass tail appears at \{$M_{\textrm{star}}=1.8 M_{\odot}$, $ R_{M_{\textrm{1.8}}} = 12~\textrm{km}$\} and, from that moment, the saturated quarks begin to take most of the total baryon density increment ($\partial n_{B}/\partial n_{\tilde{Q}} \simeq 1$). The portion of the saturated quarks at the neutron star core can be estimated from the evolution of the EoS and corresponding density profiles. For the $M_{\textrm{max.}}=2.03 M_{\odot}$ state [on the curve of $\alpha=-0.2\textrm{(w)}$ in Fig.~\ref{fig5}(b)], the portions at the neutron star core can be found as $n_{\tilde{Q}}\simeq 0.26 n_B$ and $\varepsilon_{\tilde{Q}}\simeq 0.27 \varepsilon_{\textrm{qy.}}$. For the $M_{\textrm{max.}}=2.01 M_{\odot}$ state [on the curve of $\alpha=0.2\textrm{(w)}$ in Fig.~\ref{fig5}(b)], they can be found as  $n_{\tilde{Q}}\simeq 0.33 n_B$ and $\varepsilon_{\tilde{Q}}\simeq 0.35 \varepsilon_{\textrm{qy.}}$. The scale of sound velocity appears as $\textrm{max.}[c^2_s] \simeq 0.65$ in the both cases. The resulting stellar mass number and evolution tendency are comparable with the results of Refs.~\cite{Marczenko:2020jma, Marczenko:2020wlc, Annala:2019puf, Motornenko:2019arp}, although the fundamental physical principle is different from the quarkyonic matter concept.

\section{Summary and Discussion}\label{sec4}

In this work, the single-flavor excluded-volume model~\cite{Jeong:2019lhv} is extended to the three-flavor model, including the effect of the Pauli exclusion principle. The quasifree quark states saturate dynamically the low phase space by the hard-core repulsive nature of the baryon system, which leads to multiflavor shell-like distributions of the baryons. For the application to the multiflavor system, the quark phase measure is modified from the one used in Ref.~\cite{Jeong:2019lhv}. Also, we assumed strong and weak correlation strengths among the confined quark momenta distributed around the lower boundary of the baryon shell. The pressure increases stiffly by two or three steps with emergence of the shell-like distributions, which is a different feature from the result of other works where all the quark degrees of freedom appear simultaneously~\cite{McLerran:2018hbz, Jeong:2019lhv, Zhao:2020dvu}. The sound velocity shows its peak value as $\textrm{max.}[c^2_s] \simeq 0.6$ ($\textrm{max.}[c^2_s] \simeq 0.7$) for the strong (weak) correlation assumption for the confined quark momenta. This stiff evolution ensures a hard enough EoS to support the $2M_{\odot}$ states regardless of the strangeness configuration.  The maximum mass state appears as \{$M_{\textrm{max.}}=2.03 M_{\odot}$, $ R_{M_{\textrm{max.}}}=11.4~\textrm{km}$\} under the condition of $n_0 = 5\rho_0$, $\alpha=-0.2$, and the weak correlation of the confined quark momenta.

The details of the multiflavor configuration are closely related to the correlation strength of the confined quark momenta distributed around the lower boundary of the shell-like distribution as the presence of the baryon shell increases the quasifree quark chemical potential. By following the large-$N_c$ quarkyonic matter hypothesis, one can suppose a strong correlation among the confined quark momenta where all the confined quark momenta are  distributed closely to each other. In this assumption, the flavor asymmetry of the saturated quark is unfavorable because it leads to a large enhancement of quark chemical potential:  the lower boundary of the baryon shell gets a large enhancement by the Pauli exclusion principle ($k_F^b \simeq N_c \textrm{max.} \big[ k_F^{Q_i}\big] $). Meanwhile, if one supposes a chiral symmetry restored phase in the lower boundary of the shell-like distribution, the confined state may look like the color-singlet correlated state of non-perturbative quarks. In this configuration, one can assume a weak correlation among the confined quark momenta which allows a relatively broad distribution of the confined quark momenta on the lower boundary of the baryon shell. Under this weak correlation assumption, the flavor asymmetry of the saturated quark is allowed as the quark chemical potential does not get a large enhancement by the Pauli exclusion principle ($k_F^b \leq N_c \textrm{max.} \big[ k_F^{Q_i}\big] $). The strangeness configuration can be understood by the same argument. If the repulsive core size of the $\Lambda$ hyperon is larger than the size of nucleons ($\alpha>0$), the $\Lambda$ degree of freedom is suppressed for all densities. Instead, the $s$ quarks can take a larger portion than the light quarks at the high densities, because there is no shell-like $\Lambda$ distribution whose presence increases the chemical potential of the $s$ quark. In the opposite case where $\alpha<0$, the shell-like $\Lambda$ distribution is generated and the saturated $s$ quark density does not exceed the $d$ quark density because the $d$ and $s$ quark Fermi momentum have to support the shell-like $\Lambda$ distribution as well as the shell-like nucleon distribution.

In comparison with the previous study~\cite{Duarte:2020xsp} where the stiff evolution was not evident enough, this excluded-volume model approach reproduces the required stiff evolution of EoS even for the three-flavor circumstance. The existence of the $\Lambda$ degree of freedom would be required  to support $2 M_{\odot}$ state in the high densities. At least, we demonstrated that the repulsive hard core of baryons and the dynamically generated quarkyonic-like configuration can be an alternative approach for understanding dense nuclear matter via fundamental principles. However,  more improvements are needed as the current model cannot cover all the possible range of the massive neutron stars~\cite{Cromartie:2019kug} and accommodate the matter properties at low densities. If one keeps the physical scale of the hard-core radius~\cite{Ishii:2006ec, Inoue:2016qxt, Nemura:2017bbw, Hatsuda:2018nes, Inoue:2018axd,  Sasaki:2019qnh}, various types of  potential~\cite{Hebeler:2013nza, Gandolfi:2013baa} can be referred to for the low density regime and the model can be refined to satisfy the low density matter constraints. Meanwhile, one can improve the current model in the vdW EoS framework~\cite{Hamada:1962nq, Herndon:1967zza, Carnahan:1969, Bethe:1971xm, Kurihara:1984mh, Rischke:1991ke, Kievsky:1992um, Stoks:1994wp, Wiringa:1994wb, Yen:1997rv, Machleidt:2000ge, Vovchenko:2015vxa, Zalewski:2015yea,  Redlich:2016dpb, Alba:2016hwx, Vovchenko:2017cbu, Vovchenko:2017zpj, Motornenko:2019arp, Lourenco:2019ist, Dutra:2020qsn}. For example, the baryon part of the current model can be improved as well by following the treatment of  the Carnahan-Starling modification~\cite{Carnahan:1969, Vovchenko:2017cbu, Lourenco:2019ist, Dutra:2020qsn}, where the additional repulsive contribution is reflected in the larger repulsive core size than the estimated scale in Refs.~\cite{Ishii:2006ec, Inoue:2016qxt, Nemura:2017bbw, Hatsuda:2018nes, Inoue:2018axd,  Sasaki:2019qnh}. In either approach, the required soft nature for the low densities and stiffer nature for the high densities can be achieved by additional attractive and repulsive contributions to the EoS. 

In microscopic aspects, there may be debates about the baryon-like state located on the lower boundary of the shell-like distribution. In this model, the baryon like state is clearly distinguished from the saturated quark states and the nonperturbative regulator $\Lambda_Q$ is introduced for the quark phase measure. However, depending on the nature of chiral symmetry restoration~\cite{Cohen:1991nk, Hatsuda:1992bg, Brown:2001nh, Kaiser:2007nv, Kaiser:2008qu,  Kojo:2009ha, Fukushima:2010bq} and the quark confinement mechanism around the quark Fermi surface~\cite{Ma:2019xtx, Fujimoto:2019sxg, Ma:2019ery, Rho:2020eqo}, the baryon-like state can be differently understood. One may question whether it would be still the baryon state with restored chiral symmetry or the correlated state of quarks under non-perturbative dynamics. The similarities and discrepancies between the quarkyonic matter concept and the other approaches which involve the quark degrees of freedom would be understood via further studies about the possible baryon-like states since the phase transition nature should also be related with the strong correlation patterns of quarks on the surface.

\acknowledgements
The authors acknowledge useful discussions with Larry McLerran and Sanjay Reddy during development of this work. The authors also thank Toru Kojo and Gerald Miller for inspiring discussions. Authors acknowledge the support of the Simons Foundation under the Multifarious Minds Program Grant No. 557037. The work of D.C.D., S.H.-O., and K.S.J. 
was supported by the U.S. DOE under Grant No. DE-FG02-00ER41132.

\appendix
\section{Possible emergence of $\Delta(1232)$ isobar}\label{appxa}

The $\Delta(1232)$ isobar may emerge via the energetic collisions or in  dense neutron rich matter. In this work, the low density configuration ($n_B \leq 3\rho_0$) appears as the neutron rich matter (see the profiles plotted in Fig.~\ref{fig3}). If one assumes similar size of the repulsive core for the baryons ($\omega_{n,p}=\omega_{\Delta}=1$, $n_0=5\rho_0$), the chemical potentials of baryon~\eqref{bchmp} can be written as follows:
\begin{align}
\mu_{n} &=  \left(  \frac{n_0 }{n_0- n_{n} }\right) \left( {K^n_F}^2 + m_{n}^2 \right)^{\frac{1}{2}}   -  \frac{1}{\pi^2 n_0}  \int_{0}^{ {K^n_F}} dk k^2  \left(k^2+m_{n}^2 \right)^{\frac{1}{2}},\label{bchn}\\
\mu_{p} &=m_{p}+   \frac{1}{n_0} \left\{ \bar{n}_{n}^{ex} \left(  {K^n_F}^2+ m_{n}^2 \right)^{\frac{1}{2}} -  \frac{1}{\pi^2 }  \int_{0}^{ {K^n_F}} dk k^2  \left(k^2+m_{n}^2 \right)^{\frac{1}{2}} \right\},\label{bchp}\\
\mu_{\Delta} &=m_{\Delta}+   \frac{1}{n_0} \left\{ \bar{n}_{n}^{ex} \left(  {K^n_F}^2+ m_{n}^2 \right)^{\frac{1}{2}} -  \frac{1}{\pi^2 }  \int_{0}^{ {K^n_F}} dk k^2  \left(k^2+m_{n}^2 \right)^{\frac{1}{2}} \right\},\label{bchd}
\end{align}
where $m_n=1~\textrm{GeV}$, $m_{\Delta}= 1.3~\textrm{GeV}$, and the neutron rich circumstance ($n_B\simeq n_n$, $n_p,n_{\Delta}\simeq 0$) is understood. Firstly, $\mu_{\Delta}>\mu_n$ in all of the relevant density regime  ($n_B \leq 3\rho_0$). If one considers the possible emergence via the scattering $nn\rightarrow p \Delta^{-}$,  the energy relation is satisfied in the relevant densities ($n_B \simeq 2.5\rho_0$). However, this scattering barely happens as the momentum conservation is not always matched.  Another possibility can be imagined in our model as $n+d \rightarrow \Delta^{-}+u$ after the saturation of the $d$ quark Fermi sea. In this scenario, the liberated $u$ quark falls down to the lower phases space but the emerging $\Delta^{-}$ should fill the phase space from the lower shell boundary ($k_F^{\Delta} \simeq 3 k_F^d$). Under the simplified configuration where $n_B \simeq n_n + n_{\tilde{d}}$, the baryon~\eqref{qchemp} and quark~\eqref{bchmp}  chemical potentials can be written as follows:
\begin{align}
\mu_{n} &=  \left(  \frac{n_0 }{n_0- n_{n} }\right) \left( {\left[k_F+\Delta\right]_{n}}^2 + m_{n}^2 \right)^{\frac{1}{2}}   -  \frac{1}{\pi^2 n_0}  \int_{k^{n}_F}^{ \left[k_F+\Delta\right]_{n}} dk k^2  \left(k^2+m_{n}^2 \right)^{\frac{1}{2}},\label{bchn2}\\
\mu_{\Delta} &= \left( { k_F^{\Delta} }^2 + m_{\Delta}^2 \right)^{\frac{1}{2}}  +  \frac{1}{n_0}\left\{   \bar{n}_{n}^{ex}\left(\left[k_F+\Delta \right]_{n}^2 + m_{n}^2 \right)^{\frac{1}{2}} -  \frac{1}{\pi^2} \int_{k^{n}_F}^{\left[k_F+\Delta\right]_{n}} dk k^2  \left(k^2+m_{n}^2 \right)^{\frac{1}{2}}\right\} ,\label{bchd2}\\
\mu_{d} &= 2 \left( 1- \frac{  n_{n}}{n_0} \right)   \frac{ {k_F^{n}}^2 }{{k_F^d}^2+\Lambda_d^2}   \left\{ \left( \left[k_F+\Delta\right]_{n}^2 + m_{n}^2 \right)^{\frac{1}{2}}  - \left( {k^{n}_F}^2+ m_{n}^2 \right)^{\frac{1}{2}}  \right\}+  \left(\left( k^{d}_F \right)^2 + m_{d}^2 \right)^{\frac{1}{2}},\\
\mu_{u}  &= \left( 1- \frac{  n_{n}}{n_0} \right)   \frac{{k_F^{n}}^2}{ \Lambda_u^2}   \left\{ \left( \left[k_F+\Delta\right]_{n}^2 + m_{n}^2 \right)^{\frac{1}{2}}  - \left( {k^{n}_F}^2+ m_{n}^2 \right)^{\frac{1}{2}}  \right\}+ m_{u},
\end{align}
where $k_F^n = k_F^{\Delta} =3 k_F^d$ is assumed in the small $k_F^d$ limit.  In this case, $\mu_n+\mu_d < \mu_\Delta^{-}+\mu_u$ around the expected saturation moments of the quark Fermi sea  ($ 3\rho_0 \leq n_B \leq 5\rho_0$) but the energy relation can be satisfied when the isospin asymmetry of the saturated quarks is large. However, it is unlikely to accommodate $\Delta$ isobar degrees of freedom in the quarkyonic-like system as that large flavor asymmetry of saturated quarks does not appear under the physical constraints.

\bibliographystyle{aipauth4-1}

\end{document}